\documentclass[a4paper]{article}
\usepackage{multirow}
\usepackage{subfig}
\usepackage{graphicx}
\usepackage{INTERSPEECH2022}
\usepackage{amsmath}
\usepackage{amssymb}% http://ctan.org/pkg/amssymb
\usepackage{pifont}% http://ctan.org/pkg/pifont
\usepackage{arydshln}
\usepackage{caption}

\newcommand{\cmark}{\ding{51}}%
\newcommand{\xmark}{\ding{55}}%

\title{AudioTagging Done Right: 2nd comparison of deep learning methods for environmental sound classification\vspace{-0.2cm}}
\name{Juncheng B Li, Shuhui Qu, Po-Yao Huang, Florian Metze}
\address{Carnegie Mellon University}
\email{junchenl, fmetze@cs.cmu.edu}

\begin{document}

\maketitle
\begin{abstract}
\vspace{-0.2cm}
%   This paper studies straightforward, yet must-know baselines given the recent progress in audio event detection. 
% Similar to computer vision and natural language processing, audio classification has seen tremendous progress over the last couple years. 
After its sweeping success in vision and language tasks, pure attention-based neural architectures (e.g. DeiT)~\cite{touvron2021training} are emerging to the top of audio tagging (AT) leaderboards~\cite{gong2021ast}, which seemingly obsoletes traditional convolutional neural networks (CNNs), feed-forward networks or recurrent networks. 
However, taking a closer look, there is great variability in published research, for instance, performances of models initialized with pretrained weights differ drastically from without pretraining~\cite{gong2021ast}, training time for a model varies from hours to weeks, and often, essences are hidden in seemingly trivial details. 

% In this paper, we go back to the basics, and take a step further, 
This urgently calls for a comprehensive study since our 1st comparison~\cite{li2017comparison} is half-decade old. 
In this work, we perform extensive experiments on AudioSet~\cite{gemmeke2017audio} which is the largest weakly-labeled sound event dataset available, we also did analysis based on the data quality and efficiency.
We compare a few state-of-the-art baselines on the AT task, and study the performance and efficiency of 2 major categories of neural architectures: CNN variants and attention-based variants.
We also closely examine their optimization procedures.
% Moreover, we study their robustness against natural and adversarial perturbations.
% Our intuition is that the different ways in which these models process their input (i.e. CNNs have strong locality inductive biases, which Transformers do not have) should lead to observable differences in performance and/ or robustness, an understanding of which will enable further improvements. 
Our opensourced experimental results\footnote{https://github.com/lijuncheng16/AudioTaggingDoneRight} provide insights to trade off between performance, efficiency, optimization process, for both practitioners and researchers.\footnote{This paper is under review at Interspeech 2022} 
% Our experiment results suggest that pretraining is not always necessary, our best network without pretraining could reach competetive performance compared to SOTA.
\end{abstract}
\vspace{-0.2cm}

\section{Introduction}
Recently, after seeing tremendous success in language tasks\cite{Vaswani:2017:AYN:3295222.3295349}, the ML community has been exploring a variety of methods for deploying attention-based architectures, e.g. Vision Transformers (ViT)~\cite{dosovitskiy2020image}—in computer vision and other fields, and competitive performances are reported. Recently, AST \cite{gong2021ast} and PSLA~\cite{gong2021psla} improved the SOTA performance of the AT task\footnote{Audio Tagging (AT) task aims to characterize the acoustic event of an audio stream by selecting a semantic label for it.} on the AudioSet benchmark by leveraging a suite of improvements including DeiT\cite{touvron2021training} ( distilled ViT) architecture, ImageNet pretraining, data augmentations, and ensemble.
% mainly due to the different nature of the audio input than vision or language input sequences.
However, there is still no clear ``winner-takes-all" approach in audio classification tasks that can have the best performance while being efficient.

Audio signals, 1D continuous by nature, require different processing than vision (2D) or language input sequences (discrete).
Environmental sounds, compared to well-studied human speech, do not require language model, but are more diverse and span a wide range of frequencies. Thus, the same techniques which worked well on speech are not guaranteed to work out of the box~\cite{wang2018polyphonic}. Plus, the lack of well-defined strongly-labeled data makes environment sounds recognition task not only more challenging but also understudied so far. 
% From a pure machine learning perspective, the AER task could be a better angle to gain a deeper understanding of a specific neural architecture without worrying much about the strong contextual dependencies or senses in spoken language. Therefore, we focuses on audio related tasks in the scope of this paper.

5 years ago, in our first comparison~\cite{li2017comparison}, we identified CNN's superiority over MLPs, and RNNs. CNNs~\cite{Ford2019, kong2019panns} and its variant CRNNs~\cite{wang2019comparison} have become the de-facto architecture for acoustic event recognition tasks, and have dominated the AudioTagging leaderboard until the rise of ViT~\cite{gong2021ast}. The major difference between convolutional models and attention-based networks (e.g. ViTs) are the locality inductive biases. In a nutshell, convolutional neural network sweeps through every consecutive pixel with its learned kernel, which is perfect for learning \emph{local features}\cite{li2017comparison}, but cannot see beyond its receptive field; whereas ViT networks skip through, attend in between the patches to build \emph{global correlations}, and sometimes not even rely on the positional information.

In this work, we seek to thoroughly understand the difference between using each type of the models on the AT task. We train 4 variants of transformer networks including CNN+Transformer, Vision Transformer (ViT), Transformer, and Conformer on the task of Audio Tagging, compare them with ResNet, CRNN control group. In between these 6 architectures, we perform analysis on the largest available dataset: Google AudioSet~\cite{gemmeke2017audio}.
% VGGSound~\cite{chen2021localizing} and the Kinetic Sound dataset. 
% On top of understanding the performance \& efficiency difference, we seek to gain a better understanding of how these architectures's robustness against noise perturbations to inputs and to the model parameters, and thus we could understand these models that parallels our knowledge about convolution.
% In practise, We leverage natural occlusions and adversarial examples to probe the trained models to gain deeper insights.
%  \sq{various types of no space to add adversarial}
% Our investigations provide the ML community with a better understanding of how well transformer architectures tackle audio input, the range of applications to
% which they may be deployed, and provide potential avenues for how they may be improved in terms of performance or efficiency.
Our contributions:
% Single Modality (Audio-Only) Cases:\footnote{We are not studying video-only cases since the dominant modality is audio on AudioSet, where video-only models' performances are way lower than audio-only models, making it more difficult to draw insights.}
\begin{enumerate}
    \item We systematically compared the performances of different transformer variants between several CNN variants under a permutation of different settings on the same large-scale dataset Audioset (e.g. w/. or w/o. pretraining. lr scheduling).
    % Our observations suggest that attention models bear very different \emph{model-based inductive bias} than pure convolution counterparts.
    Our experiments suggest that pretraining is \emph{not} always necessary, LR scheduling \& data augmentation are always helpful.
    % Meanwhile, models within the same category tend to behave very similarly despite the difference in other training hyper-parameters.
    \item Our experiments shed lights on critical optimization strategies \& tradeoffs, e.g. feature size, LR schedule, batch size, momentum, normalization, loss landscape, which have not been thoroughly explained previously.
    % \item We measure robustness of these different neural architectures, and we conclude that, trained on the same datasets, self-attention architectures are more robust to audio noise perturbations than convolutional models in all regards.
    % are more susceptible to \textit{patch-based} attacks whereas pure CNN models are more susceptible to \textit{high-frequency} perturbations.
    % \item We probe different architectures with natural and sythetic noise and analyze the robustness of different models.
    % \item We also measured the different architectural variants in terms of convexity, and we found out convexity of the model bears a linear correlation with the robustness of the model architecture.
\end{enumerate}
\vspace{-0.3cm}
\begin{table*}[t]
\caption{Different training procedures for single-checkpoint Audio-Only models trained on the full AudioSet as of Mar 2022. SSL\cite{srivastava2021conformer}: pretrained on Self-supervised task using 3.9M(67k hours) proprietary data. To standardize reporting difference in steps VS. epoches, epoch* here means 1 full iteration of the TrainSet. Adam optimizer's default $\beta_1$ =0.9,$\beta_2$=0.999, the ones without specified $\beta$ are default ones. LR: learning rate. TimeSpecAug\cite{Park2019SpecAugmentAS}: $t,f$ indicate max length of time and frequency mask, \cmark means specific setups are unknown. Label enhancement\cite{gong2021psla} involves altering the original labels.  CNN14~\cite{kong2019panns} achieves 0.442 mean average precision (mAP) using 128 mel bins. We are aware of PaSST~\cite{koutini2021efficient}, due to its high similarity to AST except for dropout/ensemble, we do not separately list it.} 
\hskip-0.8cm\begin{tabular}{l|cccc:cc|cc:cc}
\toprule
& \multicolumn{6}{c|}{Previous approaches}  & \multicolumn{4}{c}{Our Implementation}               \\\hline
Model & \multicolumn{4}{c|}{CNN Variants}  & \multicolumn{2}{c|}{Attention} & \multicolumn{4}{c}{see Table~\ref{tab:model-perform}}          \\\hline
\multirow{2}{*}{Procedure}        & \begin{tabular}[c]{@{}c@{}}PANNs \\CNN14\\ (\cite{kong2019panns})\end{tabular} & \multirow{2}{*}{\shortstack{ResNet\\~\cite{Ford2019}}} & \multirow{2}{*}{\shortstack{PSLA\\~\cite{gong2021psla}}}  & \multirow{2}{*}{\shortstack{ERANNs\\~\cite{verbitskiy2021eranns}} } &\multirow{2}{*}{\shortstack{Conformer\\~\cite{srivastava2021conformer}}}  & \multirow{2}{*}{\shortstack{AST\\~\cite{gong2021ast}}}  & \multirow{2}{*}{A1} & \multirow{2}{*}{A2} & \multirow{2}{*}{A3} & \multirow{2}{*}{A4} \\ \hline
params   & 42.2M  & 23M   & 13.6M     & 54.5M & 88.1M    & 88M   &  \multicolumn{4}{c}{see Table~\ref{tab:model-perform}}      \\ \hline
train dataset     & 1934187 & 1953082  & 1953082 & 1803891   & 2063949 & 1953082 & \multicolumn{2}{c:}{1998999} & \multicolumn{2}{c}{1998999}    \\
eval set          & 18887   & 19185    & 19185   & 17967     & 20371    & 19185   & \multicolumn{2}{c:}{20126}   & \multicolumn{2}{c}{20126}     \\
feature size      & \footnotesize{64$\times$1001} & \footnotesize{64$\times$1000}  & \footnotesize{128$\times$1056}  & \footnotesize{128$\times$1280} & \footnotesize{64$\times$500} & \footnotesize{128$\times$1024}  & \multicolumn{2}{c:}{128$\times$1024} & \multicolumn{2}{c}{64$\times$400}    \\ \hline
pretrained        & \xmark      & \xmark       & ImageNet  & \xmark      & SSL       & ImageNet  & ImageNet &  \xmark    &  ImageNet     & \xmark\\ \hline
epoch*             & 10    & 50       & 30        & 9    & 100     & 5         & 10       &  10       & 10        & 10        \\
batch size        & 32      & n/a      & 100       & 32     & 640       & 12        &   20       &  400       &    80      &  448      \\
optimizer         & adam    & $\underset{0.95-0.999}{\text{adam}}$ & $\underset{0.95-0.999}{\text{adam}}$ & adam & $\underset{0.9-0.98}{\text{adam}}$   & $\underset{0.95-0.999}{\text{adam}}$ & adam &  adam &   adam     & adam       \\
maxLR                & 0.001   & 0.0001  & 1.0E-04   & 0.001   & 3.0E-04    & 1.0E-05& 1.0E-05 & 4.0E-4   &   1.0E-05     &  4.0E-4       \\
LR decay          & \xmark  & \xmark   & step      & one-cycle&linear   & step      & step     & step        & step        &  step       \\
decay rate        & \xmark  & \xmark   & 0.5       & \cite{smith2019super}& 3.00E-6 & 0.5      & 0.5      &  0.5       &  0.5       &   0.5      \\
decay epochs     & \xmark  & \xmark   & 5         & cyclic             & 100       & 2        & 2        &  2       &   2      &  2       \\
weight decay      & \xmark  & 5.0E-07 & 5.0E-07   & \xmark  & 0.01     & 5.0E-07  & \xmark        &  \xmark       & \xmark        &  \xmark       \\
warmup steps      & \xmark  & \xmark   & 1000      & \xmark  & 10k      & 1000      & 1000     & 1000        &  1000       & 1000        \\
% weight averaging  & -       & -        & yes       & -       & -        & yes       &          &         &         &         \\
dropout           & \cmark  & \cmark   & \xmark    & \xmark  & 0.1      & \xmark    & \xmark        & \cmark       & \xmark     & \cmark   \\ \hline
databalancing    & \cmark  & \xmark   & \cmark    & \cmark  & \cmark   & \cmark    & \cmark   & \cmark  & \cmark  & \cmark \\
mixup             & \cmark  & \xmark   & \cmark    & modified& \cmark   & \cmark    & 0.3      &  0.3    &    0.3  &   0.3  \\
TimeSpecAug       & \cmark  & \xmark   & \cmark    & \cmark  & timeonly & \cmark    & t192,f36     & t192,f36   & t75,f12  & t75,f12        \\
label enhance     & \xmark  & \xmark   & \cmark    & \xmark  & \xmark   & \xmark    & \xmark   & \xmark  & \xmark  & \xmark \\
normalize &  \xmark  &  \xmark  & x2   &\xmark     & \xmark  & x2       & x2    & x2   &  x2 & x2        \\
train time        & 3 days  &  n/a     &  a week         &   n/a       &    n/a       & a week    & 108 Hrs         & 16.45Hrs     &  21.0Hrs  & 8.84Hrs        \\
GPU         & V100$\times$1  &  n/a    & titanX$\times$4 &    n/a      &     n/a      & titanX$\times$4& V100$\times$4        & V100$\times$4  &   V100$\times$2       & V100$\times$2  \\ \hline
Best mAP               & 0.431   & 0.392    & 0.439     & 0.450    & 0.415    & 0.448     & 0.430    &  0.437  & 0.410   & 0.411   \\\hline
\end{tabular}
\label{tab:comparison}
\vspace{-0.5cm}
\end{table*}
% \noindent \textbf{Neural Architecture Evolution on AER task:} CRNN~\cite{kong2019panns,wang2018polyphonic} has dominated the DCASE challenge lederboard for a long time. ResNet architectures were also shown to be competetive~\cite{Ford2019, verbitskiy2021eranns}. Recently, AST \cite{gong2021ast} shows ViT architectures could achieve similar performance with imageNet pre-training.  \cite{verbitskiy2021eranns} emphasized that the pure CNN network is more efficient, and could be trained several times faster than the ViT networks. Conformer based architectures~\cite{srivastava2021conformer} also shown good performance with self-supervised pretraining. Is there a winner among all these contenders? Our answer is complicated just as \cite{wightman2021resnet} has shown that ResNet could achieve competetive performance given the curated learning rate scheduling.
\section{AT Background \& Related Works}
\noindent \textbf{AudioSet~\cite{gemmeke2017audio}} contains 2,042,985 10-second YouTube video clips, summing up to 5,800 hours annotated with 527 types of sound events (weak label\footnote{does not specify which second specific event happens}). The same group~\cite{hershey2021benefit} conducted quality assessments of these labels ranging from 0-100\%.
% , and also released a set of temporally-strong labels (103,464 train, 16,996 eval). 
The \emph{full} trainset has 2 subsets: class-wise \emph{balanced} set (22,176 samples) and \emph{unbalanced} (2,042,985 samples) set, and \emph{eval} set with 20,383 samples~\cite{vggish}. \\
\noindent \textbf{AudioTagging Benchmark:}
The left columns of Table~\ref{tab:comparison} list state-of-the-art \emph{single models} and training procedures for the AT task trained on the \emph{full} AudioSet and test on the \emph{eval} set~\footnote{Previous benchmarks are excerpted from the original publications}. Here, we \emph{do not} include \emph{multi-modal} or \emph{ensembled} models, which would introduce tremendous extra variability, making fair comparison almost impossible~\footnote{We consider weight-averaging(WA)\cite{izmailov2018averaging} an implicit ensemble, hence we report single-checkpoint score w/o WA: AST~\cite{gong2021ast} 0.459$\rightarrow$ 0.448, and PSLA~\cite{gong2021psla} 0.444 $\rightarrow$ 0.439. WavegramCNN in PANNs~\cite{kong2019panns} is also an implicit ensemble, thus left out from comparison.}. Unlike some previous works, we \emph{do not} list models trained only using the \emph{balanced} subset since it only accounts for 1\% of the training set of AudioSet, which are 10-20\% mAP worse than models trained on the \emph{full} set.
% \noindent \textbf{Audio Event Recognition:} Given the difficulties for blending different modalities\cite{wang2020makes} MBT\cite{nagrani2021attention} introduced a more natural paradigm of fusing the multiple modalities and showed very good performance. 
% idea has quite close connection to image inpainting
\vspace{-0.3cm}
\section{Experiments \& Results}
We explored many variations with different optimizing techniques, data augmentations, choice of regularization, and a reasonable amount of grid search for the hyperparameters. We offer 4 different training procedures with different costs and performance that covers different typical use cases, see table~\ref{tab:comparison}.\\
\noindent \textbf{Procedure A1:} aims at reproducing the best SOTA performance. It is therefore the longest in terms of training time using the larger feature.\\
\noindent \textbf{Procedure A2:} is to test whether we could reach SOTA without pretraining, except with larger batch size and LR.\\
\noindent \textbf{Procedure A3 \& A4:} aim at matching SOTA performance using 5.12x($\frac{1}{2}$ \#Mels, $\frac{1}{2.56}$ less time resolution) smaller features thus a lot faster. It can be trained on-average 6x faster and could be a good setting for exploratory research.\\

\begin{table}[ht]
\setlength\tabcolsep{3.2pt}
\begin{center}
\caption{Our implementation of architectures with different training procedures mentioned above. A1, A3 contain blanks since ImageNet pretrained weights are only available for ResNet50 and AST/ViT. \textbf{Bolded:} the models reported in Table~\ref{tab:comparison}}.
\begin{tabular}{c c c c  | c c  } 
\toprule
\multirow{2}{*}{\textbf Model} & \multirow{2}{*}{\textbf \#param} &\multicolumn{2}{c|}{Best mAP}& \multicolumn{2}{c}{Best mAP}\\
&& {A1} & {A2}   &   {A3} & {A4}    \\
\midrule
% ViT & 0.421 & 0.969 & 2.631  & 0.193 & 0.901 & 1.961 & 0.121 & 0.841 & 1.831\\
AST/ViT &87.9M & \textbf{0.430} & 0.274 &\textbf{0.410} & 0.268 \\
Transformer & 28.5M & - & 0.230&- &0.209 \\
CNN+Trans & 12.1M & -  & \textbf{0.437} & - & \textbf{0.411}\\
Conformer & 88.1M & - & 0.335 & - &  0.308  \\
ResNet50 &25.6M & 0.410  & 0.399 & 0.382 &  0.370\\
CRNN  &10.5M & - & 0.429 & - & 0.406\\
\bottomrule
\end{tabular}
\label{tab:model-perform}
\end{center}
\vspace{-1cm}
\end{table}

\begin{figure*}[t]
    \centering
    \includegraphics[width=1.0\linewidth]{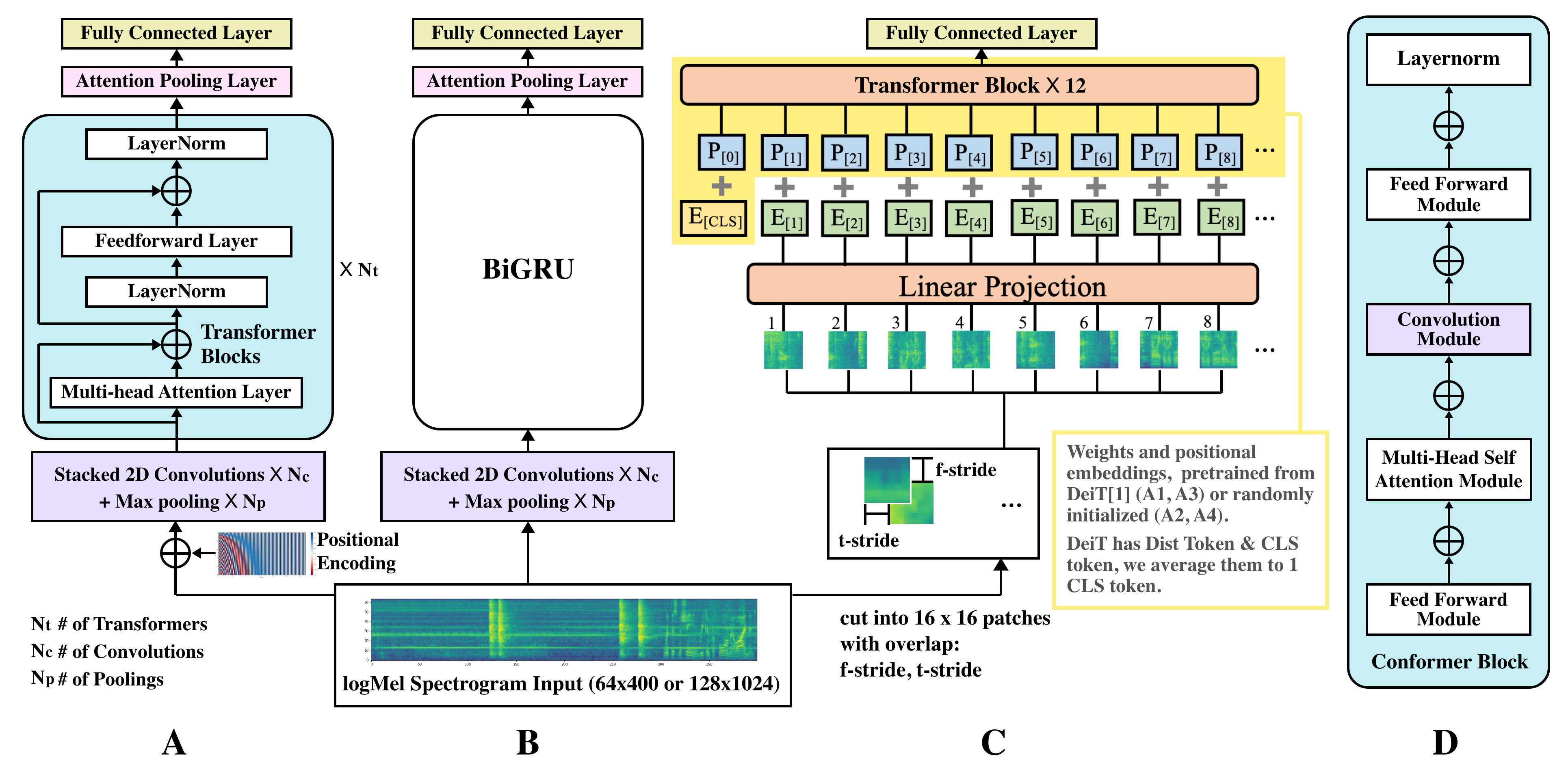}
    \caption{The overall architecture (A): CNN+Transformer (B): TALNet\cite{wang2019comparison}(CRNN) (C): AST\cite{gong2021ast}/DeiT\cite{touvron2021training}(ViT) (D): Conformer}
    \label{fig:architecture}
    \vspace{-0.5cm}
\end{figure*}
% positional encoding wrong place
\noindent \textbf{AST/ViT} (Table~\ref{tab:model-perform}, Figure~\ref{fig:architecture}(C)). uses pre-trained weights from the DeiT-base-384~\cite{touvron2021training} model imported from the timm library~\cite{wightman2021resnet}. In order to preserve the learned positional embedding in the pre-trianed DeiT~\cite{touvron2021training} model (24$\times$24=576 patches), we choose the same 16$\times$16 patch-size for all our AST experiments. We did comparison on stride sizes as shown in Table~\ref{tab:patches}. e.g. A3 procedure use strides of 8 (8 pixels overlap), resulting in (7$\times$49=343) patches. Shorter strides benefits smaller features more than larger features, but there's a catch: Computation cost grows quadratically due to the $\mathcal{O}(p^2)$ attention mechanism, $p$ being the number of patches. 
% \vspace{-0.3cm}
\begin{table}[ht]
\setlength\tabcolsep{3.2pt}
\begin{center}
\caption{time, freq stride influence AST's mAP and train time, blanks are the experiment could not fit into GPUs.}
\begin{tabular}{c|c|cccccc} 
\toprule
\tiny{Procedure}& \tiny{tstride$\times$fstride} &{4$\times$4}  &  {6$\times$6} & {8$\times$8} & {10$\times$10}  &  {12$\times$12} & {16$\times$16} \\
\midrule
\multirow{3}{*}{\shortstack[c]{A1\\ \tiny{128$\times$1024}}} & mAP &  -    &   -    &   -    & \textbf{0.430} & 0.429 & 0.421 \\
&\#patches & 8192 & 3570 & 2048 & 1212 & 756 & 512\\
&\footnotesize{Hrs/Epoch} &- & -& -& 10.8 &6.19 & 5.53\\\hline
\multirow{3}{*}{\shortstack[c]{A3\\ \tiny{64$\times$400}}} & mAP& \textbf{0.414} & 0.408 & 0.410 & 0.370  & 0.320 & 0.24\\
&\#patches & 1485 & 390 & 343 & 195 & 128 & 72\\
&\footnotesize{Hrs/Epoch} & 12.67 &2.45&2.1&1.44 & 1.01 & 0.8\\
\bottomrule
\end{tabular}
\label{tab:patches}
\end{center}
\vspace{-1cm}
\end{table}\\
\noindent \textbf{Transformers} (Table~\ref{tab:model-perform}) We implemented pure transformers without convolution or ViT type of patches, which takes in logMel Spectrogram as input. 
Since transformer layer is temporal agnostic, we introduce positional encoding to retain the temporal order of inputs. To add positional encoding to the audio model, we adopted the classic positional encoding~\cite{Vaswani:2017:AYN:3295222.3295349}, which is defined as following, where $d$ represents the dimension of the input, $pos$ is the position in time, and $i$ is the dimension index in the input tensor.\\
% \small
% \begin{equation}
$\mathbf{P} \mathbf{E}_{(p o s, k)}=\left\{\begin{array}{ll}{\sin \left(p o s / 10000^{2 i /d}\right)} & {k=2i} \\ {\cos \left(p o s / 10000^{2 i / d}\right)} & {k=2i+1}\end{array}\right.$\\
\label{eq:pos}
% \end{equation}
% \normalsize
We added positional encoding to spectrograms before feeding into CNN layers and we scale up the original input. Given the input x: $n \times d$, we scale up input by square root of the input dimension:
$x = x \cdot round\lfloor\sqrt{d}\rceil + \mathbf{PE}(x)$\\
The overall architecture is depicted in Figure~\ref{fig:architecture}(A) but without the stacked 2D conv + pooling layers.
The resulting mAP are not as high as the AST/ViT type. In Table~\ref{tab:model-perform}, we report the best performing one with $N_t=8$ layers of Transformer blocks with MultiHeadAttention (MHA) layers (8 attention heads) after testing out $N_t=2, 4, 6, 8, 10, 12$ with 4, 8, 12 attention heads setups. Same as~\cite{koutini2021efficient}, to train faster and generalize better, we implemented dropout layers within the MHA layers. mAP for other setups are too low for meaningful comparisons. Transformers suffer from the same $\mathcal{O}(n^2d)$ cost issue as ViT, $n$ being the input length. Therefore, training for Transformers is quadratically slower than models with the same number of parameters without the attention mechanism. \\
\noindent \textbf{CRNN} (Table~\ref{tab:model-perform}, Figure~\ref{fig:architecture}(B)) We follow the well-tuned TALNet architecture using $N_c=10$ convolution layers with 3$\times$3 kernels and $N_p=5$ max-pooling layers in between, resulting in embedding size of 1024 feeding into the biGRU layer. The output of GRU is fed to a fully-connected layer of size 527 to predict frame-wise probabilities. Finally, the frame probabilities is aggregated with a pooling function to make final prediction. We did not tune hyper-params for TALNet $\sim$
$\mathcal{O}(md^2) $ where $m$ is the stacked convolution output sequence length, $d$ is the representation dimension. \\
\noindent \textbf{CNN+Transformer} (Table~\ref{tab:model-perform}, Figure~\ref{fig:architecture}(A))
The first part of Stacked Conv blocks and Pooling is the same as the CRNN we implemented, which uses $N_c=10$ Conv layers and $N_p=5$ pooling layers.
We replace the GRU layer with the aforementioned Transformers\footnote{To avoid repetition, $m, n, d$ are shared among different models}$\sim\mathcal{O}(m^2d)$, and experimented with $N_t=1, 2, 4$ layers of Transformer layers with 4,8,12 attention heads. We use the same positional encoding as our aforementioned Transformer model.
Here, we report the best performing $N_t=2$ layers of transformers with 8 attention heads, followed by the attention pooling layer to aggregate frame probabilities.\\
\noindent \textbf{ResNet} (Table~\ref{tab:model-perform}) Following~\cite{Ford2019}, we replace the standard ResNet50's last layer with attention pooling layer to output 527 classes probabilities. We also implemented ResNet34, but since ResNet architectures are from the same family$\sim \mathcal{O}(knd^2)$, $k:$ kernel size, we only list ResNet50 in Table~\ref{tab:model-perform}.\\
\noindent \textbf{Conformer} (Table~\ref{tab:model-perform}) We implemented the same large conformer as~\cite{srivastava2021conformer}$\sim\mathcal{O}(n^2d)$ using 12 conformer blocks with 768D encoder embeddings and 12 attention heads, as this setting was reported to be the best performing setup for conformer architecture. Each conformer block is shown in Figure~\ref{fig:architecture}(D).
\vspace{-0.2cm}
\section{Observations and Discussion}
\begin{figure}[t]
    \hskip-0.6cm\includegraphics[width=1.2\linewidth]{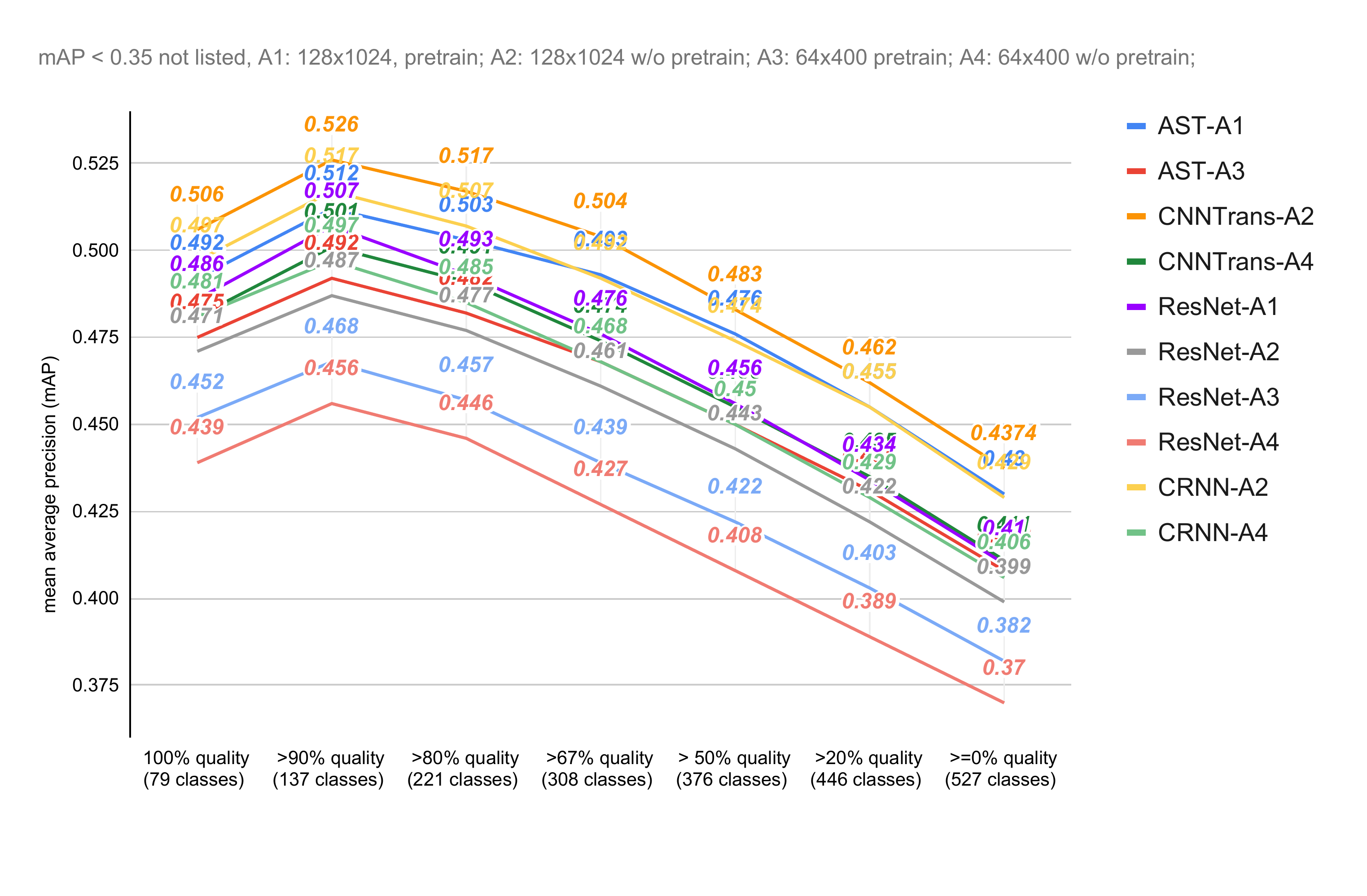}
    \vspace{-0.9cm}
    \caption{Performance(mAP) fluctuation on quantiles of data with varying label quality.}
    \label{fig:labelquality}
\vspace{-0.6cm}
\end{figure}
% \begin{table}[ht]
%     \centering
%      \caption{Performance with quantiles of data with different label quality. (number of classes in the parenthesis in 1st colum)}
%     \begin{tabular}{l|lllll}
%     \hline
%         Model & AST & AST & CNNTrans & ResNet & CRNN \\ 
%         Procedure &A1 & A3 & A4 & A4 &A4\\\hline
%         100\%(79) & 0.504 & 0.475 & 0.469 & 0.431 & 0.445 \\ 
%         $\geq$90\%(137) & 0.525 & 0.492 & 0.489 & 0.442 & 0.457 \\ 
%         $\geq$80\%(221) & 0.518 & 0.481 & 0.472 & 0.431 & 0.441 \\ 
%         $\geq$67\%(308) & 0.509 & 0.467 & 0.457 & 0.41 & 0.421 \\ 
%         $\geq$50\%(376) & 0.492 & 0.45 & 0.436 & 0.392 & 0.401 \\ 
%         $\geq$20\%(446) & 0.474 & 0.431 & 0.417 & 0.373 & 0.382 \\ 
%         $\geq$0\%(527) & 0.430 & 0.408 & 0.397 & 0.354 & 0.362 \\ \hline
%     \end{tabular}
%     \vspace{-0.5cm}
%     \label{tab:labelquality}
% \end{table}
% Table~\ref{tab:pre-train} shows the comparison of all the SOTA models on AudioSet(Audio-Only) as of Mar 2022. The bottom six models are our own implementations. As we can see, pre-trained models from other large dataset tend to prevail in terms of performance. We could achieve competetive performance on the audio-only task. 
\vspace{-0.1cm}
\subsection{Data Quality \& Data Efficiency}
\vspace{-0.1cm}
\noindent \textbf{Missing files:} In table 1, due to the downloading difference of AudioSet, number of train\&test set varies \emph{a whopping $\pm5\%$} across previous works, especially the different test size could cause severe fluctuations in final mAP reporting as seen in Figure~\ref{fig:labelquality}. e.g. one could have downloaded the lower-label-quality test samples that tank their score. We release our test set labels along with our implementation for consistency. \\
\noindent \textbf{Speedup:} As we can see in Table~\ref{tab:comparison}, previous approaches benefited from using larger features. In Table~\ref{tab:patches} we see dramatic speedup using 64(\#mels)$\times$400(time) logMel spectrogram\footnote{The waveform is downsampled to 16 kHz; frames of 1,024 samples (64 ms) are taken with a hop of 400 samples (25 ms); each frame is Hanning windowed and padded to 4,096 samples before taking the Fourier transform; the filterbank of 64 triangle filters spans a frequency range from 0 Hz to 8 kHz.} VS. 128(\#mels)$\times$1024 filter bank features, this 5.12 times feature size reduction results in more than 6 times speedup.
Note that \emph{half of} the pipeline efficiency (for both train \& inference) depends on the data efficiency, the rest is then about model efficiency. Using our smaller feature would result in $10^{-1}$ reduction of data time per sample. Therefore, we highly recommend our A3, A4 procedure for research or inference task, which could easily fit into a single 16GB-GPU with a lot less carbon footprint.
Indeed, we trade-off 2 points of mAP for 6 times of speed up. To analyze the performance loss due to feature temporal/frequency resolution loss, we took a closer look at different label quality quantiles of the data. As Figure~\ref{fig:labelquality} shows, larger features outperform smaller features in high-quality classes by 2$\pm$0.1\% mAP, indicating models benefit, but very limited, from higher freq/time feature resolution.
\vspace{-0.1cm}
\subsection{Impact of Pretraining}
Similar to the conclustion of~\cite{he2019rethinking}, we find ImageNet pretraining helps certain architectures to converge, but is not a must. As is shown in Table~\ref{tab:model-perform}, CNN-Trans model \emph{trained from scratch} can \emph{outperform} AST models that are pre-trained on ImageNet and also be better than the Conformer model pretrained using SSL in~\cite{srivastava2021conformer}. CRNNs are also competetive.
We observe AST models needs pretrained weights to converge to optimal, cold-start weights/positional embeddings from random Gaussian results in a lot lower performance. This is maybe due to huge distribution shift from Gaussian to optimal weight distribution.
\vspace{-0.2cm}
\subsection{Insights regarding Optimization Process}
\label{sec:optimization}
LR scheduling is crucial for training a high-performance model. The proven strategy is to scale \textbf{LR $\propto$ batch size}~\cite{smith2018don} when amplifying batch size, and to fill the GPU-memory till full with large enough batch size, since increasing batch-size linearly speeds up training time.
\textbf{LR decay} is also important, and it can be viewd as simulated annealing~\cite{smith2018don}. We view Adam/SGD training as  $\frac{dw}{dt}=-\frac{d\mathcal{L}}{dw}+\eta(t)$ where $\mathcal{L}$ is the BCE loss function in our case, summed over all training examples, and $w$ denotes the parameters. $\eta(t)$ denotes Gaussian random noise updating in continuous “time” $t$ towards convergence, which models the effect of estimating the gradient using mini-batches.
In~\cite{smith2018don}, they showed that the mean $\mathbb{E}(\eta(t))= 0$ and variance $\mathbb{E}(\eta(t)\eta(t'))=gF(w)\delta(t-t')$, where $F(w)$ describes the covariance in gradient fluctuations between different parameters. They also proved that the
“noise scale” $g=\epsilon(\frac{N}{B}-1)$, where $\epsilon$ is the learning rate (LR), $N$ the training set size and $B$ the batch size. This noise scale controls the magnitude of the random fluctuations in the training dynamics.
Intuitively, the initial noisy phase allows the model to explore a larger fraction of the parameter space without getting trapped in local minima. Once we find a promising region of parameter space, we reduce the noise to fine-tune the parameters. In Table~\ref{tab:comparison} and all our experiments, we observe LR decay outperforms constant LR by at least 2 \% mAP. We also implemented cyclic schedule~\cite{smith2019super} used by \cite{verbitskiy2021eranns}, but the result is suboptimal VS. step decay. Meanwhile, annealing the temperature in a series of discrete steps can sometimes trap the system in a “robust” minimum, in our experiments, this severely impede training attention-based models such as AST, transformers. e.g when we initialize training AST in our A3 procedure with a non-optimal LR: say 1E-6, 1E-4, the model's mAP get ``stuck" below 0.1. This effect is less severe for non-attention or hybrid models. We postulate the culprit is the \emph{sharp loss landscape} of the attention-based models~\cite{chen2022when}, when the gradient accumulation cannot adapt to changes in the loss landscape, training would be impeded since we would keep on exploring the wrong direction in parameter space. This also explained why our training also failed when we used \emph{gradient accumulation} step size$\geq 2$.
In Table~\ref{tab:comparison}, we can see some previous works have tuned the $\beta_1$ of Adam to 0.95, this is actually letting gradient updating \textbf{twice} as slower than the default 0.9. $m_t = \beta_1 m_{t-1}+(1-\beta_1) g_t$ where $m_t$ is the first moment of the gradient and $g_t$ is the gradient at $t$ step. This trick might compensate for the aforementioned sharp-loss-landscape phenomenon, but we did not achieve the same performance gain using the trick, hence stick to default.
Some previous works advocate \emph{regularization} for better model generalizability by adopting weight decay or dropout, we observe 1-2\% mAP performance drop using regularization in our experiments, and therefore did not include them in our procedures.
We note ~\cite{koutini2021efficient} using AdamW~\cite{loshchilov2017decoupled} optimizer, whose purpose is to perform weight decay within the adam update iteration, hence is not studied here.
\vspace{-0.2cm}
\subsection{Data Augmentations \& Normalization}
AST and PSLA \cite{gong2021ast, gong2021psla} applied an unusual normalization (Normalize*) to the feature input: $x = \frac{x-\mu}{2\sigma}$, which results in the normalized feature with $\mathbb{E}(x) = 0$, and $var(x)=0.25$.
In our experiment, Training AST using input of $var(x)=1$ or $var(x)=0.0625$ would lead to mAP of 0.09, 0.02 respectively. The latter phenomenon can be explained by vanishing gradients, whereas the former behavior is likely due to the training noise $\eta(t)$ (\S\ref{sec:optimization}) being too large, when input variance is too large, the training would keep on exploring suboptimal regions of the parameter space.  
Previous works in Table~\ref{tab:comparison} reported that data augmentation tricks improve performance across all the models/procedures. Table~\ref{tab:augment} shows their specific impact after ablation. Note that Mixup, TimeSpecAug shift the input distribution's mean and variance, the higher the mixup coefficient/the longer time/freq mask$\rightarrow$the lower variance of the input$\rightarrow$more influence on end performance. We find data balancing helps, but due to the stochasticity(random sampling/shuffling) of the vanilla (no-balancing) train loader, its influence also fluctuates.
\begin{table}[h]
\vspace{-0.5cm}
\setlength\tabcolsep{2.0pt}
\begin{center}
\caption{Data Augmentation's influence on mAP through ablation, on all models trained with A4 procedure}
\begin{tabular}{c|ccccccc} 
\toprule
\tiny{Augmentation}& dataBalancing & Mixup &  SpecAug & Normalize* &   \\
\midrule
mAP drop(\%) & 3.9$\pm$2.0 & 1.5$\pm$0.8 & 1.5$\pm$0.3 & 16.5$\pm$9.0 \\
\bottomrule
\end{tabular}
\label{tab:augment}
\end{center}
\vspace{-0.8cm}
\end{table}\\
\noindent \textbf{Takeaways}:
Smaller features: 6$\times$ efficiency VS. 2\% mAP loss.\\
Models with Local+Global info $\rightarrow$ efficient \& best Performance
Attention-based models: More \# params, harder/more sensitive to train, sharp loss landscape, but more robust to noise\cite{li2021audio}

\bibliographystyle{IEEEtran}

\bibliography{egbib}

\end{document}